\documentclass[letterpaper,12pt]{article}   
\usepackage{osajnl2} 
\usepackage[draft]{hyperref} 
\usepackage{amssymb}
\usepackage{amsmath}
\usepackage{epstopdf}

\begin{document}

\title{Quantum repeaters with entangled coherent states}

\author{Nicolas Sangouard$^{1,2},$ Christoph Simon$^{1,3},$ Nicolas Gisin$^{1}$, Julien Laurat$^{4},$ Rosa Tualle-Brouri$^{5},$ Philippe Grangier$^{5}$}
\address{$^{1}$Group of Applied Physics-Optics, University of Geneva, CH-1211 Geneva 4, Switzerland}
\address{$^{2}$Mat\'{e}riaux et Ph\`{e}nom\`{e}nes Quantiques, CNRS UMR 7162, Universit\'e Paris Diderot, 75013 Paris, France}
\address{$^{3}$Institute for Quantum Information Science and Department of Physics and Astronomy, University of Calgary, Calgary T2N 1N4, Alberta, Canada}
\address{$^{4}$ Laboratoire Kastler Brossel, Universit\'e Pierre et Marie Curie, Ecole Normale Sup\'erieure, CNRS, Case 74, 4 place Jussieu, 75252 Paris Cedex 05, France}
\address{$^{5}$ Laboratoire Charles Fabry de l'Institut d'Optique, CNRS UMR 8501, Universit\'e Paris Sud, 91127 Palaiseau, France}

\begin{abstract}
\textit{
Entangled coherent states can be prepared remotely by
subtracting non-locally a single photon from two quantum
superpositions of coherent states, the so-called
"Schr\"odinger's cat" state. Such entanglement can further
be distributed over longer distances by successive
entanglement swapping operations using linear optics and
photon-number resolving detectors. The aim of this
paper is to evaluate the performance of this approach
to quantum repeaters for long distance quantum communications.  
Despite many attractive features at
first sight, we show that, when using state-of-the-art
photon counters and quantum memories,  they do not achieve
higher entanglement generation rates than repeaters based
on single-photon entanglement.  We discuss potential
developments which may take better advantage of the
richness of entanglement based on continuous variables,
including in particular efficient parity measurements.
}
\end{abstract}


\section{Introduction}
The distribution of entanglement over long distances is of
great interest both for fundamental tests on quantum
correlations and for applications, e.g. in the frame of
long distance quantum communication. However, it is
extremely challenging due to losses. In classical
communication, they can be overcome by amplifiers
along the transmission lines, but the no-cloning theorem
forbids such an amplification for quantum states. Quantum
repeaters \cite{Briegel98} are a potential solution to
distribute entanglement over long distances.
They require heralded distribution and storage of
entanglement within elementary links of moderate-distance
and successive entanglement swapping operations between the
links.

Single photon entanglement $\frac{1}{\sqrt{2}}(|01\rangle+|10\rangle)$ (where
$|0\rangle$ and $|1\rangle$ label either Fock states with
zero and one photon or orthogonal polarizations of single
photons) can be distributed efficiently over long distances
using either atomic ensembles or single atoms in
high-finesse cavities as quantum memories. The simplest
approach is based on atomic ensembles and linear optics
(see \cite{Sangouardreview} for a review), as initially proposed by Duan et al. \cite{Duan01} and further improved by several groups
\cite{attheo, Sangouard07}. These works have stimulated a
large amount of experiments \cite{atexp}. An alternative
approach uses individual quantum systems such as NV centers
in diamond \cite{Childress06}, spin in quantum dots
\cite{SimonNiquet07,VanLoock08} or trapped ions
\cite{Sangouard09}. Recently it was shown that the
distribution rates of entanglement offered by repeaters
based on single ions \cite{Sangouard09} are significantly
higher than those achieved with atomic ensemble based
schemes. The main reason is that entanglement swapping
operations can be performed deterministically for trapped
ions \cite{Riebe08}. In contrast, the success probability
for entanglement swapping is bounded by 1/2 for schemes
using single photons and Bell measurements based on linear
optics \cite{Calsamiglia01}.

Another approach, mostly unexplored for quantum repeaters, is using quantum continuous variables, i.e. quadrature operators for a quantized light mode \cite{bookCV}. Non classical states, such as squeezed states, two-mode squezed light or Schr\"odinger's cat states, are well known to be very fragile and thus cannot be propagated on long distances. However, it has been shown very recently how entanglement can be remotely prepared \cite{Ourjoumtsev09}. Following this line, we will consider entanglement of the form \cite{Sanders92}
\begin{eqnarray}
\label{eq1}
&&|\phi_-\rangle_{AB}=\frac{1}{\sqrt{M_-}}(|\alpha\rangle_A|\alpha\rangle_B-
|-\alpha\rangle_A|-\alpha\rangle_B), \\
&& M_-=2(1-e^{-4|\alpha|^2})\nonumber
\end{eqnarray}
where $|\alpha\rangle$ and $|-\alpha\rangle$ correspond to
coherent states with opposite phases. The subscripts $A$
and $B$ label spatial modes located at two distant
locations. Such an entangled state can be prepared remotely
by subtracting a single photon from either even
\begin{eqnarray}
&& |+\rangle=\frac{1}{\sqrt{N_+}}\left(|\alpha\rangle+|-\alpha\rangle\right) \\
&& N_+=2(1+e^{-2|\alpha|^2}) \nonumber
\end{eqnarray}
or odd superposition of Fock states
\begin{eqnarray}
\label{oddsuperposition}
&& |-\rangle=\frac{1}{\sqrt{N_-}}(|\alpha\rangle-|-\alpha\rangle) \\
&& N_-=2(1- e^{-2|\alpha|^2}) \nonumber
\end{eqnarray}
prepared at $A$ and $B$  \cite{Ourjoumtsev09}. The four
states
\begin{eqnarray}
&&|\phi_\pm\rangle_{AB}=\frac{1}{\sqrt{M_\pm}}(|\alpha\rangle_A|\alpha\rangle_B\pm
|-\alpha\rangle_A|-\alpha\rangle_B), \\
&&|\psi_\pm\rangle_{AB}=\frac{1}{\sqrt{M_\pm}}(|\alpha\rangle_A|-\alpha\rangle_B\pm
|\alpha\rangle_A|-\alpha\rangle_B), \\
&& M_\pm=2(1\pm e^{-4|\alpha|^2})\nonumber
\end{eqnarray}
can be almost perfectly distinguished in the limit of large
$\alpha$ using linear optics and photon-counting detectors
\cite{vanEnk01}. Note that they define an orthonormal
basis only in the limit of large $\alpha$ since
$\langle\psi_+|\phi_+\rangle=\frac{1}{\cosh{2|\alpha|^2}}$,
thus they are sometimes called quasi-Bell states
\cite{Hirota01}. The principle of the Bell state
discrimination is as follows : After passing through a
50-50 beamsplitter, the four Bell states become
\begin{eqnarray}
&&\nonumber |\phi_+\rangle \rightarrow |\hbox{even}\rangle_{\text{out1}}|0\rangle_{\text{out2}}, \\
&&\nonumber |\phi_-\rangle \rightarrow |\text{odd}\rangle_{\text{out1}}|0\rangle_{\text{out2}}, \\
&&\nonumber |\psi_+\rangle \rightarrow |0\rangle_{\text{out1}}|\text{even}\rangle_{\text{out2}}, \\
&&\nonumber |\psi_-\rangle \rightarrow |0\rangle_{\text{out1}}|\text{odd}\rangle_{\text{out2}},
\end{eqnarray}
where $|\text{even}\rangle_{\text{out1}}$
($|\text{odd}\rangle_{\text{out1}}$) means that the output
$1$ contains an even (odd) number of photons. By setting
one photon-number resolving detector for each output $1$
and $2$ to perform number parity measurement, one can
discriminate the four Bell states. Note however that for
small amplitude $\alpha \rightarrow 0,$ there is non-zero
probability of failure when both of the detectors do not
register a photon due to the non-zero overlap $|\langle
0|\text{even}\rangle|^2=2e^{-4|\alpha|^2}/(1+e^{-4|\alpha|^2}).$
Essentially deterministic entanglement swapping operations
are thus possible, at least for large enough amplitude
coherent states.

Remote entanglement creation and entanglement swapping are two basic
elements of a quantum repeater. It is thus natural to wonder what is the performance
of such an architecture. As said above, recent experiments offer additional motivations. Coherent-state superposition have been created with small \cite{Ourjoumtsev06} and larger amplitudes \cite{Ourjoumtsev07, Takahashi08}.
Furthermore, they have already been produced with sufficiently narrow spectral bandwidth to fulfill the
storage requirements within atomic ensembles \cite{Neergaard06}. Entangled coherent states have also been created based on non-local single-photon subtraction \cite{Ourjoumtsev09}. It has also been shown that one- or two- photon subtraction can enhance the entanglement of gaussian entangled states, by making them non-gaussian \cite{Dakna,Ourjoumtsev07b, Takahashi 09}.
Moreover quantum repeater with entangled coherent states
might be attractive : being rather simple
with the use of atomic ensembles, linear optics and
single-photon detectors, it seems efficient, a priori,
since entanglement swapping operations can be performed
deterministically.

Here, we present a detailed study of the performance of
such a quantum repeater protocol, and compare it to that of
repeaters based on single-photon entanglement with the same
ingredients. Surprisingly, we find that, when using
state-of-the-art photon counters and quantum memories,
the achievable distribution rate of coherent-state
entanglement is not higher than the one for single-photon
based entanglement. While the distribution of entangled
coherent states can be efficient within kilometer-long
elementary links, the entanglement swapping operations
require photon-counting detectors with very high
efficiency, much beyond those available to date, to swap
entanglement with high fidelity. This is particularly true
for moderate $\alpha \approx 1$ to large $\alpha > 1$
amplitude coherent states. In this case, entanglement
purification operations \cite{Jeong05} might be used but
the complexity of the repeater architecture increases
significantly in this case. Thus, if one wants to
distribute high-fidelity entangled states with relatively
simple architectures, one has to consider coherent states
with small amplitudes $\alpha \ll 1.$ In this case, the
dominant errors are due to vacuum components, and they do
not affect the fidelity of the distributed state if one
uses simple postselection techniques. However, in the limit
of small $\alpha \rightarrow 0,$ coherent state
entanglement can no more be swapped deterministically using
linear optics and photon-counting detectors leading to
distribution rates comparable to those achievable for
single-photon protocols.

In the next section, we focus on the elementary link of
this quantum repeater where entanglement is created
non-locally. In agreement to the results mentioned in Ref.
\cite{Ourjoumtsev09}, we show that the fidelity of the
entangled state and the probability to create it can be
made independent of the coherent state amplitude. In
section \ref{section3}, we consider the entanglement
swapping operations between elementary links. We compare
the efficiency of the distribution of entangled coherent
states with the one achievable for entangled photons.
Section \ref{section4} is devoted to potential
developments.

\section{Elementary link}
\label{section2}
We begin our investigation of this quantum repeater architecture by describing the elementary link.

\begin{figure}[ht]
\centerline{\includegraphics[scale=0.5]{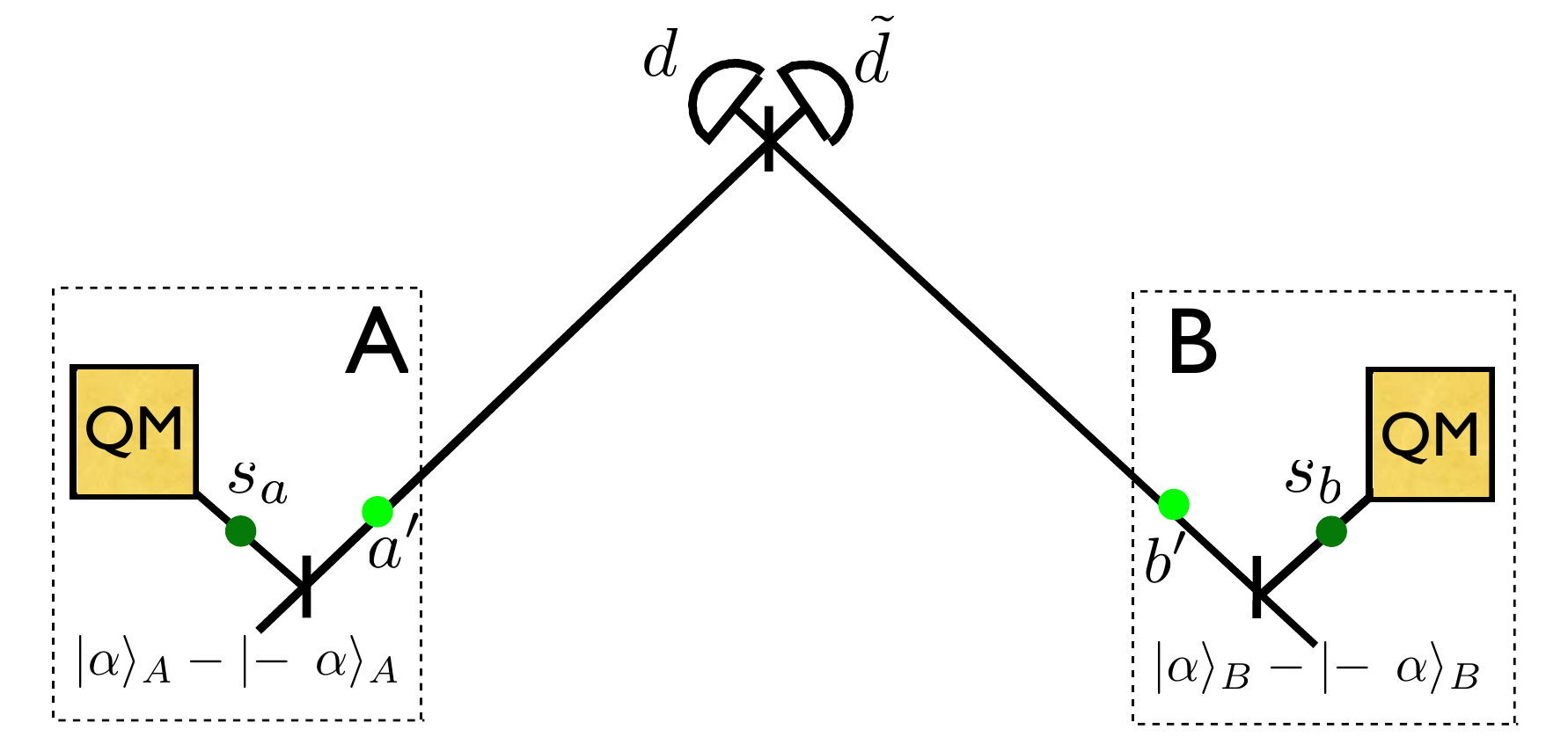}}
\caption{Schematic architecture of an
elementary link connecting two distant locations $A$ and $B$.
Memories and detectors are represented by squares and
half-circles respectively. Vertical bars denote
beamsplitters. Odd superposition of Fock states are
generated at each location and sent through an asymmetric
beamsplitter with small transmission. The reflected light
is stored in local memories whereas the transmitted part is
combined on a 50/50 beamsplitter. The detection of a single
photon at the central station heralds the storage of an
entangled coherent state $|\phi_-^\theta\rangle_{AB}$
defined in Eq. (\ref{eq3}) with high probability, due to
the asymmetry of the local beamsplitters.} \label{fig1}
\end{figure}

\textit{Principle :} Entanglement of coherent states
(\ref{eq1}) is known to be extremely sensitive to losses \cite{vanEnk01}.  However, a protocol
based on non-local photon subtraction allows one to create
such entanglement of remote modes with high fidelity even in the case of a very lossy channel
\cite{Ourjoumtsev09}. The required
architecture is presented in Fig. \ref{fig1}. It is very
similar to the one of Ref. \cite{Sangouard07} used to
entangle remote atomic ensembles with single-photon
sources.

Let us consider two distant nodes, say A and B as before, where odd superposition $$|-\rangle_A=\frac{1}{\sqrt{N_-}}(|\alpha\rangle_A-|-\alpha\rangle_A)=\frac{e^{-\frac{|\alpha|^2}{2}}}{\sqrt{N_-}}(e^{\alpha a^\dagger}-e^{-\alpha a^\dagger})|0\rangle$$ and $$|-\rangle_B=\frac{1}{\sqrt{N_-}}(|\alpha\rangle_B-|-\alpha\rangle_B)=\frac{e^{-\frac{|\alpha|^2}{2}}}{\sqrt{N_-}}(e^{\alpha b^\dagger}-e^{-\alpha b^\dagger})|0\rangle$$ have been prepared. $a$ and $b$ are bosonic operators associated to the locations A and B and $|0\rangle$ refers to vacuum state. Each state is sent through a beamsplitter with a small transmission coefficient $\sin^2\theta$, called in the following 'tapped fraction', such that $a=\cos \theta s_a + \sin \theta a'$ and $b=\cos \theta s_b + \sin \theta b'.$ The reflected light associated to the modes $s_a$ and $s_b$ are stored locally in quantum memories and the transmitted parts corresponding to the modes $a'$ and $b'$ are combined on a 50-50 beamsplitter at a central station located half-way between the two nodes. 
Omitting for simplicity the phase acquired by the photons during their transmission, the modes after the beamsplitter are $d=\frac{1}{\sqrt{2}}(a'+b')$ and $\tilde{d}=\frac{1}{\sqrt{2}}(a'-b').$ The detection of a single photon after the beamsplitter at the central station, e.g. in mode $d=\frac{1}{\sqrt{2}}(a'+b')$ heralds the storage of remote entanglement of the form\\

\begin{eqnarray}
\label{eq3}
&|\phi_-^\theta\rangle_{AB}&=\frac{1}{\sqrt{M_-^\theta}}\Big(|\alpha\cos\theta\rangle_A|\alpha\cos\theta\rangle_B-|-\alpha\cos\theta\rangle_A|-\alpha\cos\theta\rangle_B\Big),\nonumber \\
&M_-^\theta=&2(1-e^{-4|\alpha|^2\cos^2\theta})
\end{eqnarray}
 within memories located at $A$ and $B.$ Note that the conditional state associated to the detection in mode $\tilde{d}=\frac{1}{\sqrt{2}}(a'-b')$ is
\begin{eqnarray}
&|\psi_-^\theta\rangle_{AB}&=\frac{1}{\sqrt{M_-^\theta}}\Big(|\alpha\cos\theta\rangle_A|-\alpha\cos\theta\rangle_B-|\alpha\cos\theta\rangle_A|-\alpha\cos\theta\rangle_B\Big).
\end{eqnarray}
Note also that $|\phi_-\rangle$ and $|\psi_-\rangle$ have exactly one ebit of entanglement independently of the size of $\alpha$ \cite{Hirota01}. (This appears clearly if one writes $|\phi_-\rangle$ and $|\psi_-\rangle$ in the basis $\{|+\rangle,|-\rangle\}$.) This is not the case for $|\phi_+\rangle$ and $|\psi_+\rangle$ which reduce to vacuum states in the limit $\alpha \rightarrow 0$ and thus have manifestly less than one ebit of entanglement for small $\alpha$. Local one-qubit operations could thus be used to transform $|\phi_-\rangle$ into $|\psi_-\rangle$ and $|\phi_+\rangle$ into $|\psi_+\rangle.$ However the transformation, e.g. from $|\phi_-\rangle$ to $|\phi_+\rangle$ can be realized using local operations in the limit of large $\alpha$ only. The required one-qubit rotations, so far considered out of experimental reach, might be performed using the proposal of Ref. \cite{Ourjoumtsev09} which is based on quantum teleportation. In what follows, we consider that one-qubit rotations are performed in a deterministic way and that they do not introduce errors. \\

\textit{Imperfections :} Depending on the tapped fraction $\sin^2 \theta$ of the local beamsplitters, two photons can be transmitted toward the central station. Due to losses, there is a high probability for one of them to be lost and the state resulting from the detection of one photon, e.g. in mode $d,$ is orthogonal to the expected state (\ref{eq3}), i.e.
\begin{eqnarray}
\label{eq4}
&|\phi_+^\theta\rangle_{AB}&=\frac{1}{\sqrt{M_+^\theta}}\Big(|\alpha\cos\theta\rangle_A|\alpha\cos\theta\rangle_B+|-\alpha\cos\theta\rangle_A|-\alpha\cos\theta\rangle_B\Big),\nonumber \\
&M_+^\theta=&2(1+e^{-4|\alpha|^2\cos^2\theta}).
\end{eqnarray}
From a perturbative calculation $(|\alpha|^2 \sin^2{\theta} \ll 1)$, one can show that starting from odd superposition $|-\rangle,$ the state conditioned to the detection of one photon in mode $d$
\begin{eqnarray}
\label{eq5}
&&\rho_0^{AB}= \frac{1}{M_-^\theta + 2 M_+^\theta |\alpha|^2 \sin^2\theta} \times \\
&&\Big(M_-^\theta |\phi_-^\theta\rangle_{AB \-\ AB}\langle\phi_-^\theta|\nonumber+2 M_+^\theta |\alpha|^2 \sin^2\theta|\phi_+^\theta\rangle_{AB \-\ AB}\langle\phi_+^\theta|\Big)\nonumber
\end{eqnarray}
is created with the probability
\begin{eqnarray}
\label{P0}
&P_0^{\phi_-}=&\frac{1}{N_-^2}2\sin^2\theta |\alpha|^2 e^{-2\sin^2\theta |\alpha|^2} \left(M_-^\theta + 2 M_+^\theta |\alpha|^2 \sin^2\theta\right)\eta_t\eta_d
\end{eqnarray}
where $\eta_t=e^{-L_0/(2 L_{att})}$ is the fiber
transmission with the attenuation length $L_{att}$ (we use
$L_{att}= 22$ km, corresponding to losses of 0.2 dB/km,
which are currently achievable at a wavelength of 1.5
$\mu$m), $L_0$ being the distance between the nodes $A$ and
$B.$ $\eta_d$ is the detection efficiency. Since the
detection of one photon in mode $\tilde{d}$ combined with
the appropriate one-qubit rotation also collapses the state
stored into the remote memories into
$|\phi_-^\theta\rangle_{AB}$ in the ideal case, the overall
success probability $P_0$ for the entanglement creation
within an elementary link takes a factor 2, such that
$P_0=2 P_0^{\phi_-}.$ We assume single-mode quantum
memories so that one can only make a single entanglement
generation attempt per communication time $L_0/c$ where $c
= 2 \times 10^8$ m/s is the photon velocity in the fiber
\cite{Simon07}. (This supposes that odd superposition of
Fock states (Eq. \ref{oddsuperposition}) can be prepared
with a repetition rate greater than $c/L_0.$) The average
time to create the conditional state (\ref{eq5}) is thus
given by
\begin{equation}
T_{0}=\frac{L_0}{c}\frac{1}{P_0}.
\end{equation}

\begin{figure}
\centerline{\includegraphics[scale=0.4]{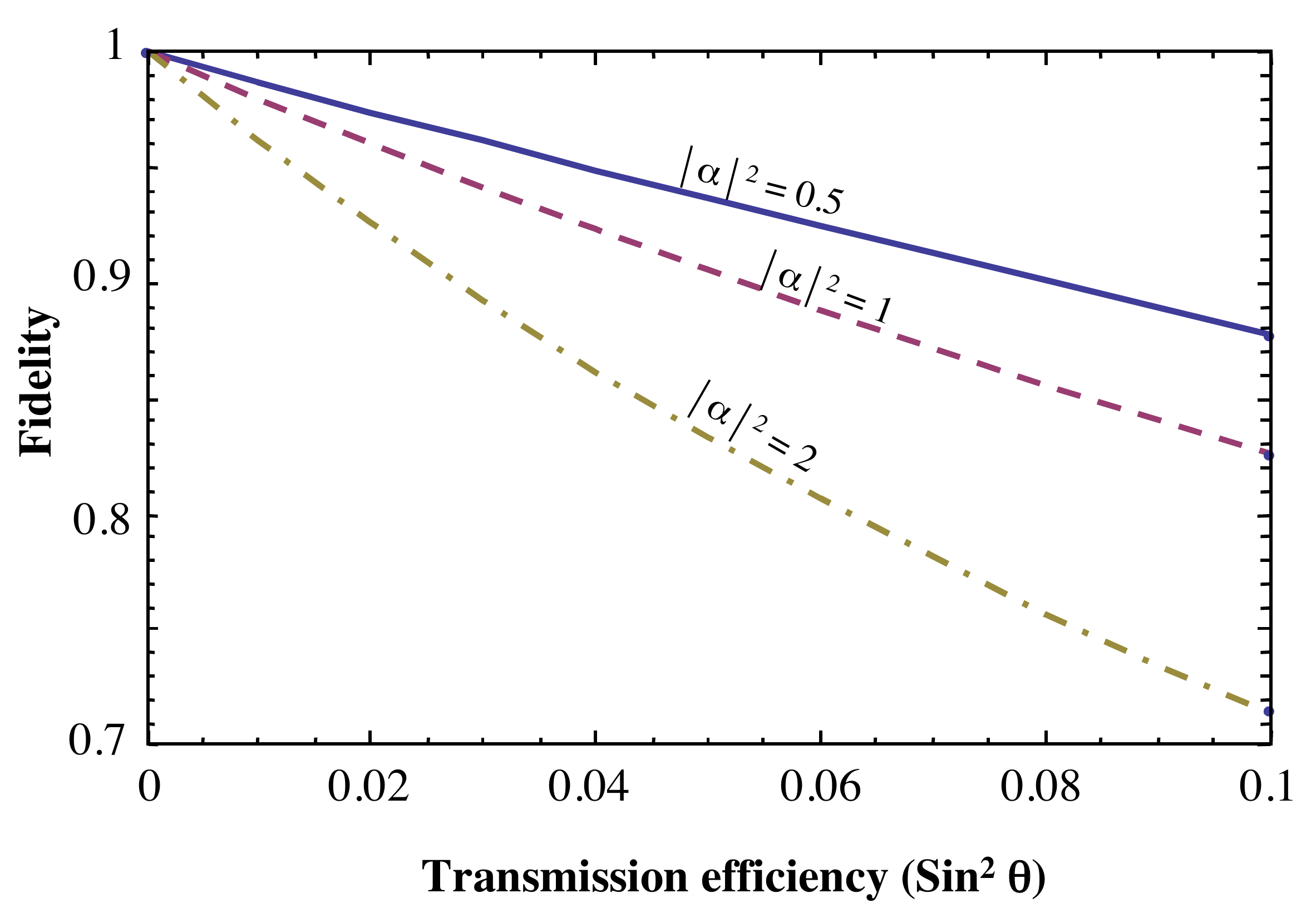}}
\caption{Fidelity of the distributed state within an elementary link versus the tapped fraction $\sin^2\theta$ of local beamsplitters. Different amplitudes $\alpha$ have been considered : $|\alpha|^2=0.5$ full line, $|\alpha|^2=1$ dashed line, $|\alpha|^2=2$ dash-dotted line. ($\eta_d$=0.9,$ L_0$=100km)}
\label{fig2}
\end{figure}

\begin{figure}
\centerline{\includegraphics[scale=0.4]{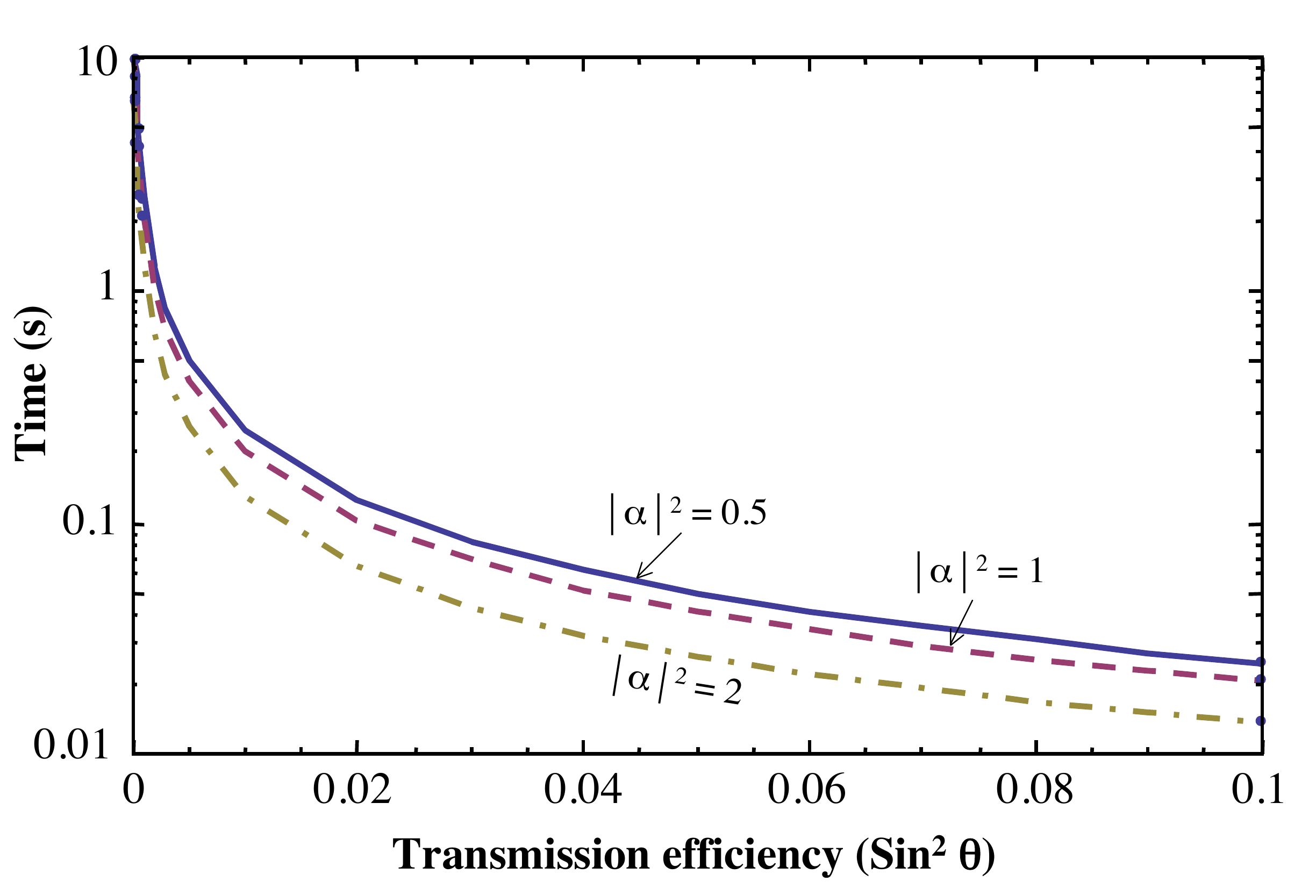}}
\caption{Average time for the distribution
of an entangled state within an elementary link versus the
transmission coefficient $\sin^2\theta$. The distance is
$L_0=$100 km and the photon-number resolving detector efficiency
$\eta_d$ is 90\%. Entangled coherent
states with different amplitudes $\alpha$ have been
considered : $|\alpha|^2=0.5$ full line, $|\alpha|^2=1$
dashed line, $|\alpha|^2=2$ dash-dotted line.} \label{fig3}
\end{figure}

\textit{Results :} In Fig. \ref{fig2}, we plot the fidelity $F_-^0$ of the conditional state
\begin{equation}
\label{fidelitylink}
F_-^0= _{AB}\langle \phi_-^\theta|\rho_0^{AB}|\phi_-^\theta\rangle_{AB}
\end{equation}
and in \ref{fig3} the average time $T_{0}$ for entanglement creation within the elementary link as a function of the tapped fraction $\sin^2 \theta$ for various amplitudes $\alpha$. The detection efficiency is set to $\eta_d=0.9$ and we consider a typical distance of $L_0=100$km. \\

\textit{Comments :}
Note that for large enough initial state amplitude and small enough transmission coefficient such that $|\alpha|^2 \cos^2 \theta \gg 1,$ the coefficients $N_-$ and $M_\pm^\theta $ are constant. This leads to $$P_0=\sin^2 \theta |\alpha|^2 e^{-2\sin^2\theta |\alpha|^2}(2+4|\alpha|^2\sin^2\theta)\eta_t \eta_d$$ and $$F_-^0=\frac{1}{1+2|\alpha|^2\sin^2{\theta}}.$$ In this regime, the entanglement distribution rate and the fidelity of the distributed state can thus be made independent of $\alpha$ by controlling the local beamsplitter transmission such that the product $|\alpha|^2 \sin^2\theta$ is constant. For example, by choosing $|\alpha|^2 \sin^2\theta \sim 0.05,$ the average time for the remote creation of an entangled pair over 100 km is of $T_0 \sim 54$ ms and the fidelity of the distributed state is of $F_-^0 \sim 90\%$ using high-efficiency photon detectors $\eta_d=0.9.$ (This is true as long as $|\alpha|^2>2$ which is sufficient to fulfill the requirement $|\alpha|^2 \cos^2 \theta \gg 1$.) This remarkable property, which has already been mentioned in Ref. \cite{Ourjoumtsev09}, reveals the interest of the setup to entangle coherent states with arbitrary amplitudes at a distance.\\

Note also that, for small amplitudes $|\alpha|^2<1$, the average time $T_{0}$ for the distribution of an entangled pair is longer when one starts with even superposition $|+\rangle.$ This can be intuitively understood since for small $|\alpha|^2,$ one can approximate $|-\rangle$ by $|1\rangle$ and $|+\rangle$ by $ |0\rangle.$ If one starts with vacuum states, there is obviously no chance to get the desired photon detection at the central station. In the opposite case where there is a single-photon at each remote location,
the studied scheme is similar to the one of Ref. \cite{Sangouard07} using single-photon sources. For $\alpha \rightarrow 0,$ the achievable distribution rate in the elementary link is thus similar to the one calculated in Ref. \cite{Sangouard07}.

\section{Entanglement swapping}
\label{section3}

 Entanglement can further be distributed over longer distances using successive entanglement swapping as proposed in Ref. \cite{vanEnk01}. We evaluate in this section the fidelity of the distributed state after swapping operations, distingushing the cases of large and small amplitude coherent states.

\begin{figure}
\centerline{\includegraphics[scale=0.5]{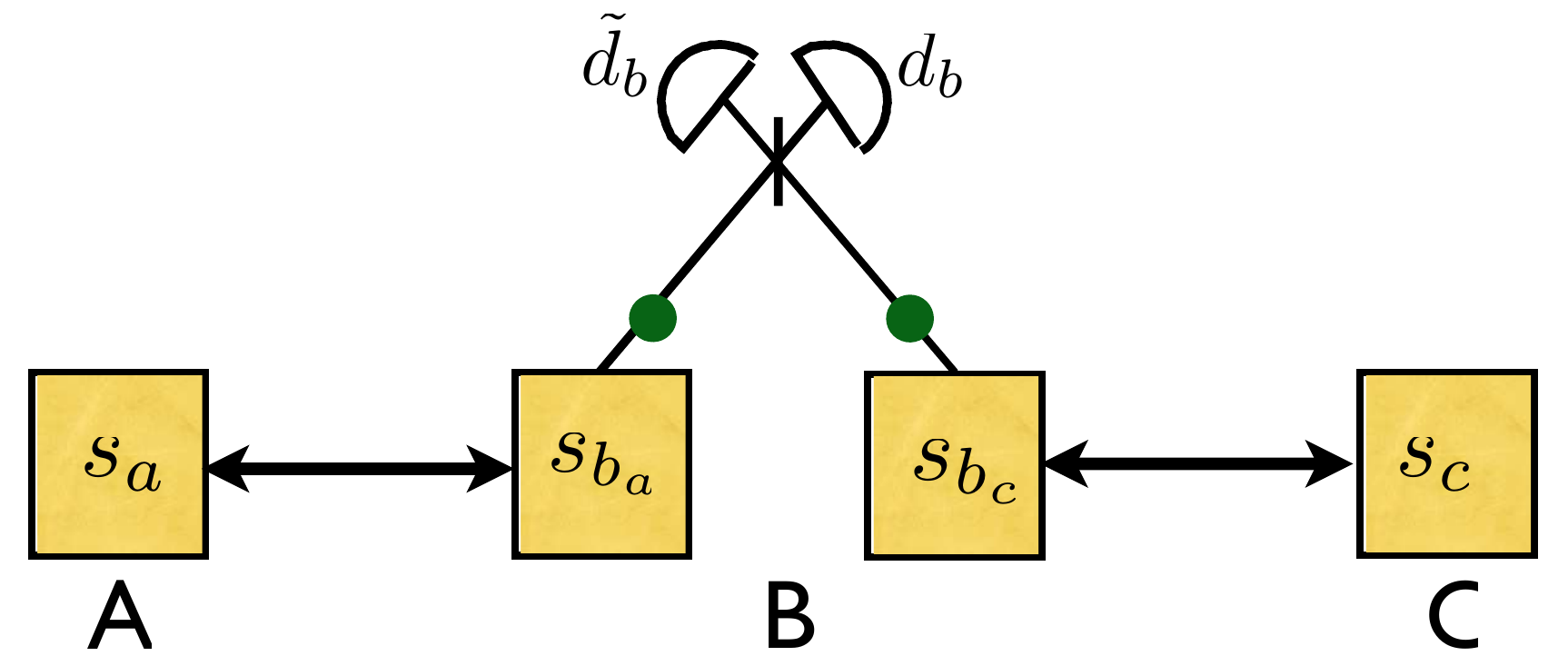}}
\caption{Entanglement swapping : Starting
from entangled links $A$-$B$ and $B$-$C,$ entanglement between $A$ and $C$ is obtained if one retrieves the light
fields stored in the memories located at $B$ location,
combines them in 50-50 beamsplitter and then counts the
photon number in the output modes $d_b$ and $\tilde{d}_b.$
Memories and detectors are represented by squares and
half-circles respectively. The vertical bar symbolizes a
beamsplitter.} \label{fig4}
\end{figure}

\textit{Principle :} Let us consider two links, say A-B and B-C, where $|\phi_-\rangle_{AB_a}$ and $|\phi_-\rangle_{B_cC}$ have been distributed and stored into memories located at A-B and B-C respectively (see Fig. \ref{fig4}). The light fields stored in mode $s_{b_a}$ and $s_{b_c}$ within the memories located at the same node B are retrieved and combined on a 50-50 beamsplitter. We then count the photons in each output modes of this beamsplitter, say $d_b=\frac{1}{\sqrt{2}}(s_{b_a}+s_{b_c})$ and $\tilde{d_b}=\frac{1}{\sqrt{2}}(s_{b_a}-s_{b_c}).$ As described in the introduction, there is no photon in one of the output modes whereas in the other, one has to determine the parity of the photon number. The memories located at locations $A$ and $C$ are projected into a specific Bell state depending on the parity measurement : if the photon number in the output mode $d_b$ $(\tilde{d_b})$ is odd, the memories A and C are projected on $|\phi_-\rangle_{AC}$ ($|\psi_-\rangle_{AC}$) whereas if it is even, the memories are in state $|\phi_+\rangle_{AC}$ ($|\psi_+\rangle_{AC}$). In principle, entanglement swapping can thus be performed almost
deterministically. However, in the limit of small amplitudes, there is non-zero probability of failure when both of the detectors do not register a photon. For $\alpha \rightarrow 0,$ the probability for the entanglement swapping operation reduces to $1/2$ for photon detections with unit efficiency as in the case of single-photon entanglement. \\

\textit{Imperfections :} Consider the first entanglement swapping operation allowing one to distribute entanglement between remote nodes $A$ and $C,$ starting from entanglement $\rho_{0}^{AB} \otimes \rho_{0}^{BC}$ between locations $A$-$B$ and $B$-$C$ with
\begin{equation}
\label{rho0}
\rho_{0}^{AB}=F_-^{0} |\phi_-\rangle_{AB \-\ AB}\langle \phi_-|+F_+^{0} |\phi_+\rangle_{AB \-\ AB}\langle \phi_+|
\end{equation}
where $F_-^{0}$ is given by Eq. (\ref{fidelitylink}) and
$F_+^{0}=1-F_-^{0}.$ Note that, depending on whether the
photon has been detected in mode $d$ or $\tilde{d}$ in the
elementary link, the heralded state can also be a mixture
between $|\psi_-\rangle_{AB}$ and $|\psi_+\rangle_{AB}$
with weights $F_-^{0}$ and $F_+^{0}$ respectively. Here, we
consider that the appropriate one-qubit unitary operation
has been applied to transform one admixture into the other
such that one starts with the state (\ref{rho0}).  We now
take into account non-unit efficiencies of quantum memories
and of single-photon detectors. We calculate the
probability for a successful entanglement swapping and the
fidelity of the conditional state, first for large
amplitude coherent states ($|\alpha|^2 \geqslant 1$), and
then concentrating on small amplitude ones ($|\alpha|^2 <
1$).

\subsection{Large amplitude coherent states}
\label{section3a} Let us first focus on the distribution of
large amplitude entangled coherent states. While the
success probability depends on the number $n$ of detected
photons at location $B$, the fidelity only depends on the
parity of $n.$

One shows that for odd $n,$ the state is
ideally $|\phi_-\rangle_{AC}$ (or $|\psi_-\rangle_{AC}$
depending on whether the light field is in mode $d$ or
$\tilde{d}$) but due to imperfections its fidelity is
reduced to $F_-^{1}=_{AC}\langle
\phi_-^\theta|\rho_{1}^{AC}|\phi_-^\theta\rangle_{AC}(=_{AC}\langle
\psi_-^\theta|\rho_{1}^{AC}|\psi_-^\theta\rangle_{AC})=\frac{N^{odd}}{D^{odd}}$
with
\begin{eqnarray}
\nonumber N^{odd}&=&\left((F_-^0)^2+\left(F_+^0\frac{M_-^\theta}{M_+^\theta}\right)^2\right)\times\\
&\nonumber \sum_{k=0}^{+\infty} & \frac{2^{2k}}{(2k)!}(1-\eta)^{2k}|\alpha|^{4k}\cos^{4k}\theta\\
\label{eq10} &+& 2F_-^0 F_+^0 \frac{M_-^\theta}{M_+^\theta} \times \\
&\nonumber \sum_{k'=0}^{+\infty}&\frac{2^{2k'+1}}{(2k'+1)!} (1-\eta)^{2k'+1} |\alpha|^{2(2k'+1)} \cos^{2(2k'+1)}\theta \\
\nonumber &=& \left((F_-^0)^2+\left(F_+^0\frac{M_-^\theta}{M_+^\theta}\right)^2\right)\times\\
&\nonumber& \cosh\left({2(1-\eta)|\alpha|^{2}\cos^{2}\theta}\right)\\
\nonumber&+& 2F_-^0 F_+^0 \frac{M_-^\theta}{M_+^\theta} \sinh\left({2(1-\eta)|\alpha|^{2}\cos^{2}\theta}\right)\-\ ;
\end{eqnarray}
\begin{eqnarray}
\label{eq11}
D^{odd}&=&\left((F_-^0)^2+2  F_-^0 F_+^0+\left(F_+^0\frac{M_-^\theta}{M_+^\theta}\right)^2\right)\times \\
&\nonumber  \sum_{k=0}^{+\infty} & \frac{2^{2k}}{(2k)!}(1-\eta)^{2k}|\alpha|^{4k}\cos^{4k}\theta\\
&&+\left((F_+^0)^2+2F_-^0F_+^0+\left(F_-^0\frac{M_+^\theta}{M_-^\theta}\right)^2\right)\frac{M_-^\theta}{M_+^\theta} \times \nonumber \\
&\sum_{k'=0}^{+\infty}&\frac{2^{2k'+1}}{(2k'+1)!} (1-\eta)^{2k'+1} |\alpha|^{2(2k'+1)} \cos^{2(2k'+1)}\theta
\nonumber \\
\nonumber &=& \left((F_-^0)^2+2  F_-^0 F_+^0+\left(F_+^0\frac{M_-^\theta}{M_+^\theta}\right)^2\right)\times \\
&\nonumber& \cosh\left({2(1-\eta)|\alpha|^{2}\cos^{2}\theta}\right)\\
&&+\left((F_+^0)^2+2F_-^0F_+^0+\left(F_-^0\frac{M_+^\theta}{M_-^\theta}\right)^2\right)\frac{M_-^\theta}{M_+^\theta} \times \nonumber \\
&\nonumber& \sinh\left({2(1-\eta)|\alpha|^{2}\cos^{2}\theta}\right).
\end{eqnarray}
Here $\eta$ is the product of the memory and detector
efficiencies $\eta=\eta_m\eta_d.$ The probability to detect
$n$ photons at the first entanglement swapping level, $n$
being odd, is given by
\begin{equation}
\label{P1odd}
P_{1}^{n, odd}=\frac{2}{M_-^\theta}\eta^n \frac{2^n|\alpha|^{2n}\cos^{2n}\theta}{n!}e^{-2|\alpha|^2\cos^2\theta} \times D^{odd}.
\end{equation}

For even $n,$ the state is ideally $|\phi_+\rangle_{AC}$
(or $|\psi_+\rangle_{AC}$ depending on whether the light
field is in mode $d$ or $\tilde{d}$) but due to
imperfections its fidelity is reduced to
$G_+^{1}=_{AC}\langle
\phi_+^\theta|\rho_{1}^{AC}|\phi_+^\theta\rangle_{AC}(=_{AC}\langle
\psi_+^\theta|\rho_{1}^{AC}|\psi_+^\theta\rangle_{AC})=\frac{N^{even}}{D^{even}}$
where $N^{even}$ and $D^{even}$ can be obtained from
(\ref{eq10}) and (\ref{eq11}) respectively by changing
$F_\pm^0$ by $F_\mp^0$ and $M_\pm^\theta$ by
$M_\mp^\theta.$ (Note that in the limit of small $\alpha$
$|\phi_+^\theta\rangle$ and $|\psi_+^\theta\rangle$ reduce
to vacuum states and $G_+^{1}$ is no longer a meaningful
fidelity.) One also shows that
\begin{equation}
P_{1}^{n, even}=\frac{2}{M_+^\theta}\eta^n \frac{2^n|\alpha|^{2n}\cos^{2n}\theta}{n!}e^{-2|\alpha|^2\cos^2\theta} \times D^{even}.
\end{equation}

\begin{figure}[ht!]
\centerline{\includegraphics[scale=0.4]{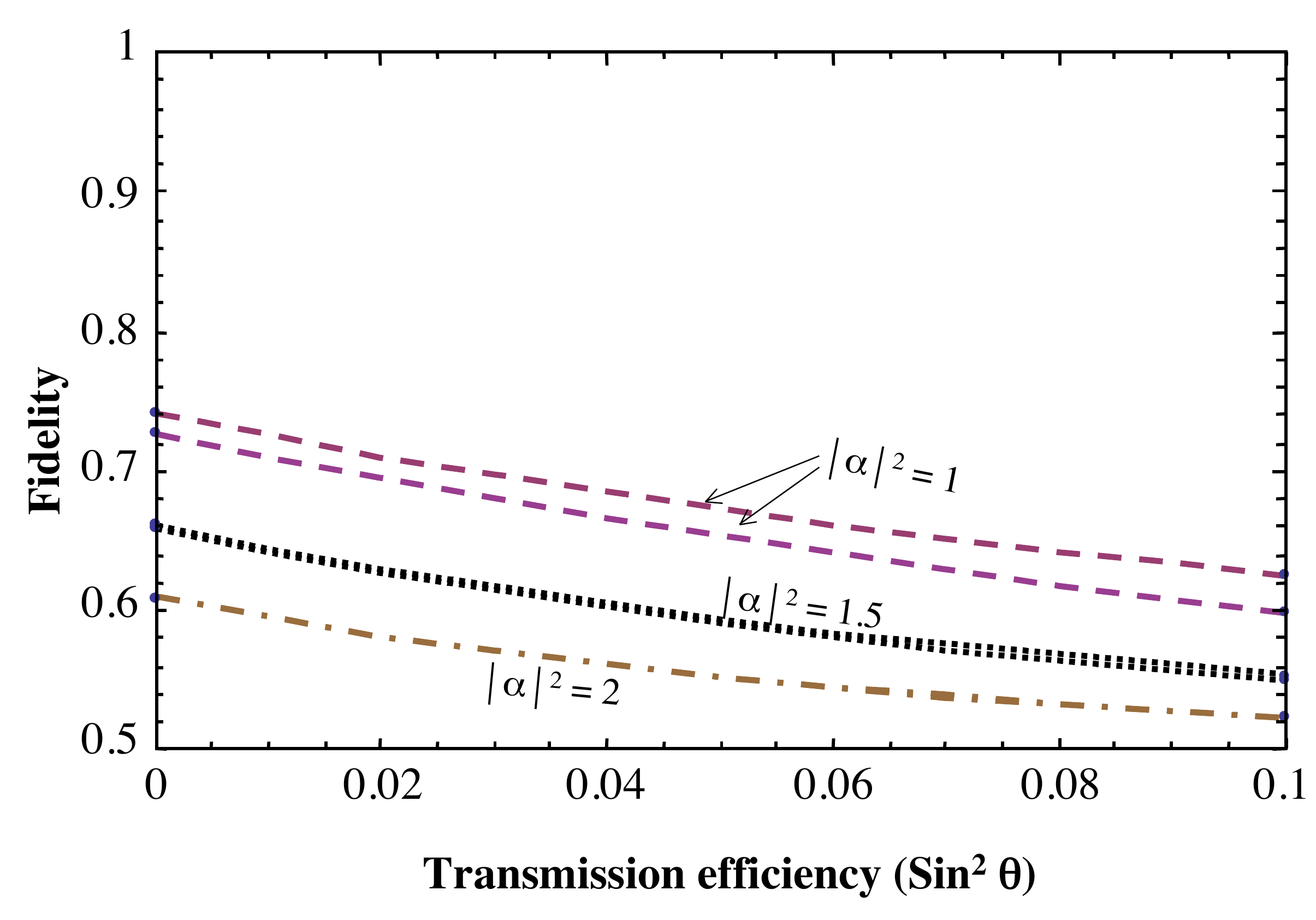}}
\caption{(Color online)
Fidelity of the conditional state after one entanglement swapping operation as a function of the tapped fraction $\sin^2\theta$ of local beamsplitters in the case of large amplitude coherent states. For coherent states with an amplitude such that $|\alpha|^2 \approx 1$, the fidelity of the distributed state depends on the parity of the number of photons detected. Memory and photon-detector efficiencies are taken equal to $\eta_m=\eta_d=0.9.$}
\label{fig5}
\end{figure}

The previous formulas can be simplified if one focuses on small tapped fraction of local beamsplitters $(\cos{\theta}^2 \approx 1),$ i.e. on small errors $F_+^0 \ll1$ at the entanglement creation level. In this case, the fidelity of the distributed state after the first swapping takes the simpler form
\begin{eqnarray}
\label{Fmoins1}
&F_{-}^{1} \rightarrow &\frac{1}{1+\tanh(2(1-\eta)|\alpha|^2\cos^2\theta) \frac{M_+^\theta}{M_-^\theta}} + o(F_+^0) \-\ \quad \-\ \\
\label{Fplus1}
&G_{+}^{1} \rightarrow & \frac{1}{1+\tanh(2(1-\eta)|\alpha|^2\cos^2\theta) \frac{M_-^\theta}{M_+^\theta}} + o(F_+^0)
\end{eqnarray}
at the lower order with respect to the error $F_+^0.$ In the limit of very large $\alpha$ where the normalization coefficients $M_\pm^\theta$ become equal, the fidelity of the conditional state after the swapping operation does not depend anymore on the parity of the number of photons detected
\begin{equation}
\label{fid1}
F_{-}^{1} \equiv G_{+}^{1} \rightarrow \frac{1}{1+\tanh(2(1-\eta)|\alpha|^2\cos^2\theta)}.
\end{equation}
In this regime, the success probability for the first entanglement swapping is given by
\begin{eqnarray}
&P_1&=\sum_{n=1}^{+\infty} \left(P_{1}^{2n-1, odd} + P_{1}^{2n, even}\right)  \\
&&\nonumber= 1-e^{-2\eta|\alpha|^2 \cos^2 \theta}
\end{eqnarray}
which tends to $1$ for high enough detection and memory efficiencies. (Note that the summation starts from $1$ because one cannot distinguish the case where both of the output modes $d_b$ and $\tilde{d_b}$ are empty and the case where either the memory or the detector fails.) In Fig. \ref{fig5}, we plot the fidelity of the state resulting from the first entanglement swapping for various large amplitudes $|\alpha| \geqslant 1$ assuming high detector $(\eta_d=0.9)$ and memory $(\eta_m=0.9)$ efficiencies.

For $|\alpha|^2=1,$ the normalization coefficients
$M_\pm^\theta$ are not equal and the fidelity of the
distributed state depends on the parity of the number of
photons detected at the location $B$ (Eq. (\ref{Fmoins1})
and (\ref{Fplus1})). For larger amplitudes, the fidelity,
which does not depend anymore on this parity, is well
approximated by Eq. (\ref{fid1}) and it rapidly decreases
when the amplitude of coherent states increases (see Fig.
\ref{fig5}). This is intuitive since larger amplitude
coherent states have larger numbers of photons in average
requiring higher memory and detector efficiencies to store
and to detect all of them. 

The main conclusion coming from
the plots of the figure (\ref{fig5}) is that even with
optimistic assumptions on memory and detector efficiencies,
entanglement swapping of large amplitude entangled coherent
states cannot lead to the distribution of high fidelity
entanglement. For example, If one wants to perform the
first swapping operation on coherent states with amplitude
$|\alpha|^2=2$ such that the fidelity of the resulting
entangled coherent is of 90\%, one would need memory and
detector efficiencies beyond 99\%. One might think to use
entanglement purification operations after each swapping
operations, for example the protocol of Ref. \cite
{Jeong05} based on linear optics and homodyne detections
where one starts with two copies of partially entangled
states and one obtains a single copy of a more entangled
state. This protocol which is probabilistic (but heralded)
purifies bit flips. In our case, phase errors dominate but
one might transform phase errors into bit flips using
one-qubit rotations. In principle, the protocol of Ref.
\cite{Jeong05} improves an initial fidelity $F_{in}$ to
$F_{out}=\frac{F_{in}^2}{F_{in}^2+(1-F_{in}^2)}.$ For large
amplitude $| \alpha |^2 >1,$ the fidelity of the
distributed state after the first swapping operation is
below 70\% such that one would need at least two
purification operations to get one entangled state with a
fidelity of 90\%. So many purification operations would
increase the complexity  and decrease the achievable
distribution rate significantly. This is not an attractive
option if one focuses on the most immediate goal to
outperform the direct transmission of photons through an
optical fiber with a rather simple quantum repeater
protocol, at least as simple as the ones based on
single-photon entanglement \cite{Duan01,Sangouardreview}
\subsection{Small amplitude coherent states}
\label{section3b}

\begin{figure}
\centerline{\includegraphics[scale=0.4]{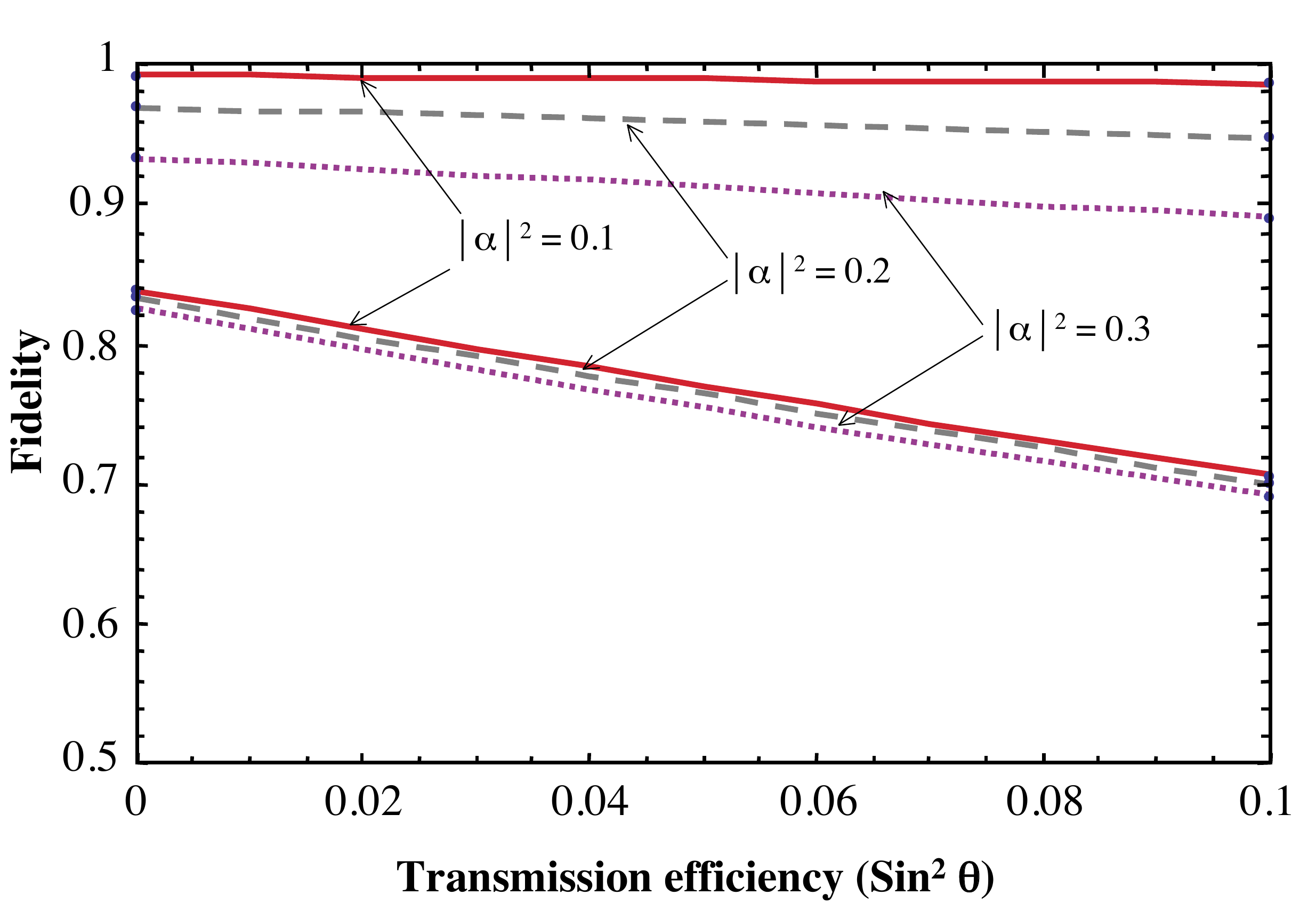}}
\caption{(Color online) Fidelity of the conditional state after one entanglement swapping operation as a function of the tapped fraction of the local beamsplitters $\sin^2\theta$ in the case of small amplitude coherent states without (three lower curves) and with (three upper curves) postselection.}
\label{fig6}
\end{figure}

We now focus on the opposite regime of small $\alpha$ and we study simpler architectures without purification steps. For small $\alpha$, the states $|\phi_+\rangle_{AC}$ and $|\psi_+\rangle_{AC}$ mainly reduce to vacuum. Since one wants to distribute entanglement rather than vacuum states, we only take into account the cases where an odd number of photons is detected at location $B$ which project ideally the memories into the state $|\phi_-\rangle_{AC}$ (or $|\psi_-\rangle_{AC}$ depending on whether the light field is in mode $d$ or $\tilde{d}$). Due to imperfections, the fidelity reduces to $F_-^{1}=_{AC}\langle \phi_-^\theta|\rho_{1}^{AC}|\phi_-^\theta\rangle_{AC}(=_{AC}\langle \psi_-^\theta|\rho_{1}^{AC}|\psi_-^\theta\rangle_{AC})=\frac{N^{odd}}{D^{odd}}$ where $N^{odd}$ and $D^{odd}$ are given by Eq. (\ref{eq10}) and (\ref{eq11}) respectively, as in the regime of large $\alpha$. The success probability for the detection of $n$ photons with $n$ an odd number, at the first entanglement swapping level is given by Eq. (\ref{P1odd}). As seen before, in the limit of small tapped fraction at local beamsplitters $(\cos^2{\theta} \approx 1)$ leading to small errors $F_+^0 \ll1$ at the entanglement creation level, the fidelity takes the simple form (\ref{Fmoins1}) which even reduces in the regime of small $\alpha$ to
\begin{equation}
F_{-}^{1} \rightarrow \frac{1}{1+(1-\eta)}.
\end{equation}
In the limit of small $\alpha,$ the entanglement swapping is mainly heralded by one-photon detections such that its success probability (in the limit of small errors $F_+^0 \ll1$) is
\begin{equation}
P_1 \approx P_{1}^{1, odd} \rightarrow \frac{1}{2}\eta(2-\eta).
\end{equation}
(As in the case of large amplitude coherent states, we do
not take into account the contribution of empty mode). The
success probability for entanglement swapping thus reduces
to one half for $\eta \rightarrow 1$ as in the case of
swapping operation for single-photon entanglement based on
linear optics. If one considers memory and detector
efficiencies of 90\% the conditional fidelity is at most of
$F_-^{odd} \approx 84\%$ after the first swapping
operations for small enough amplitude coherent states (see
the three lower curves in Fig. \ref{fig6}). This is already
quite low. However, the errors, i.e. the components
$|\phi_+\rangle_{AC}$ or $|\psi_+\rangle_{AC}$ mainly
reduce to vacuum states for small $\alpha$. One can thus
use a simple postselection technique to increase the
fidelity of the distributed state as in the case of
single-photon entanglement \cite{Duan01}.

The postselection procedure is sketched in Figure 7 in the case of a repeater with $2$ links. Each location contains two memories, denoting $A_1$ and $A_2$ for location A, and $C_1$ and $C_2$ for location C. Entangled states are established between $A_1$ and $C_1$ and between $A_2$ and $C_2.$ We are interested in the case of small $\alpha$ where these states are well approximated by states of the form $|1\rangle_{A1}|0\rangle_{A2}\pm|0\rangle_{A1}|1\rangle_{A2}.$ By converting the memory excitation back into photons and counting the number of photons in each locations $A$ and $C,$ one postselects the case where there is one excitation in each location, i.e. one at location $A$ and an other at location $C$. One thus generates an effective entanglement of the form $|1\rangle_{A1}|1\rangle_{C2}+|1\rangle_{A2}|1\rangle_{C1}.$ The vacuum component does not contribute to this final state, since if one of the two pairs of memories contains no excitation, it is impossible to detect one excitation in each location. The vacuum components thus have no impact on the fidelity of the final state, leading to the possibility to distribute entangled coherent states with small amplitudes over longer distances with high fidelity. The three upper curves in Fig. \ref{fig6} take into accounts the final postselection. One can see that for very small amplitudes, typically $|\alpha|^2 \leqslant 0.1,$ one postselects a state with almost perfect fidelity (at least not limited by empty components). For larger amplitudes, e.g. $|\alpha|^2 = 0.3,$ multi-photon components are not fully negligible but the fidelity of postselected state is higher than 90\%.\\

\begin{figure}
\centerline{\includegraphics[scale=0.5]{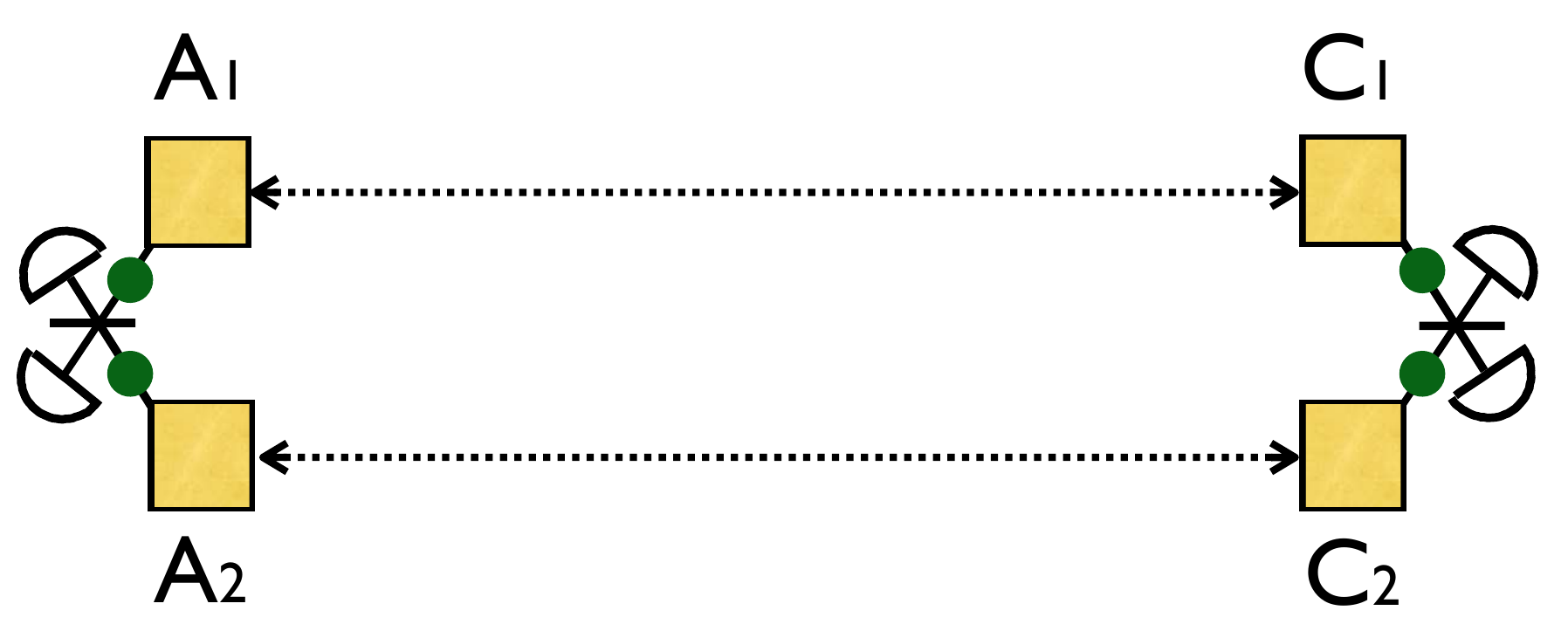}}
\caption{Schematic representation of postselection protocol. Entanglement has been distributed independently within two chains (labelled by the subscripts 1 and 2) such that the ensembles $A_1-C_1$ and $A_2-C_2$ store entangled coherent states. Light fields stored at the same location are retrieved and combined into a beamsplitter. The detection of a single-photon at each location postselects a two-photon entangled state. Memories and detectors are represented by squares and half-circles respectively. Vertical bars denote beamsplitters.}
\label{fig7}
\end{figure}

\textit{Results :} One can give examples of the typical distribution rates achievable for entangled coherent states with small amplitudes. Consider a quantum repeater with four links. (This is the optimal link number for the typical distances where the distribution rate of entanglement becomes higher than the one from the direct transmission). The average time for the distribution of an entangled state is well approximated by
\begin{equation}
T \approx \left(\frac{3}{2}\right)^2 T_0 \frac{1}{P_1 \times P_2 \times P_{\text{ps}}}.
\end{equation}
$P_1$ and $P_2$ are the success probabilities for the first
and the second swapping operation respectively and
$P_{\text{ps}}$ is the success probability for the
postselection. (see Ref. \cite{Sangouardreview} appendix A
for the factors 3/2). Note that $P_2$ can be deduced from
Eq. (\ref{P1odd}) where one has to replace $F_\mp^0$ by
$F_\mp^1$ in the expression of $D^{\text{odd}}.$ The
success probability for the postselection requires the
knowledge of the fidelity $F_-^2$ of the state resulting
from the second swapping. $F_-^2$ is analogue to $F_-^1$
where one has to replace $F_\mp^0$ by $F_\mp^1$ in the
expressions of $N^{\text{odd}}$ (Eq. (\ref{eq10})) and of
$D^{\text{odd}}$ (Eq. (\ref{eq11})). At the lowest order in
$|\alpha|^2,$ $P_{\text{ps}}$ is given by
$P_{\text{ps}}=\frac{\eta^2}{2} (F_-^2)^2.$ For given
memory and photon-detector efficiencies
$(\eta_m=\eta_d=0.9),$ we calculate the fidelity of the
postselected state versus the transmission coefficient of
local beamsplitters $(\sin^2\theta).$ This fidelity
strongly decreases when the amplitude of the initial state
increases as seen before. If one wants to distribute
entanglement with a fidelity of 0.9, one has to prepare
superposition of coherent states with small amplitudes,
typically $|\alpha|^2<0.2$ for transmission coefficient
$\sin^2 \theta < 20\%.$ We then optimized with respect to
$\alpha$ and $\theta$ the average time for the distribution
of an entangled pair for a distance $L=4 L_0=600$ km. One
finds $T=23$s for $|\alpha|^2=0.13$ and
$\sin^2\theta=0.16.$ This gives an improvement of roughly
10\% as compared to the distribution of single-photon
entanglement using the same architecture. Note however,
that the fidelity decreases of 10\% in the same time. This
is because for small but non-zero entangled coherent
states, multi-photon errors increases the success
probability for entanglement swapping but decreases the
fidelity of the postselected state. Note also that for such
distance, the average time to distribute an entangled pair
of single-photon using the direct transmission though an
optical fiber is of the order of 100s (considering a photon-pair
source with a repetition rate of $10$ GHz).

\section{Discussion and conclusion}
\label{section4}

Quantum repeaters based on entangled
coherent states seemed attractive at first sight, because
entanglement creation at a distance based on single-photon
detection allows one to distribute entanglement with high
fidelity despite losses. Furthermore, entanglement swapping
operations are deterministic, leading one to expect, a priori, distribution rates of entanglement higher than the rates that are achievable with similar schemes based on
single-photon entanglement where entanglement swapping
operations are performed with linear optics with at most
50\% of efficiency. 

Our study shows that entanglement of coherent
states with arbitrary amplitudes can be created efficiently
within tens of kilometer-long elementary links. However, it turns out from our calculations that
large amplitude entangled coherent states $|\alpha|^2
\geqslant 1$ cannot be distributed with high fidelity over
longer distances using entanglement swapping operations.
This would require either photon counting detectors and
quantum memories with unrealistic efficiencies or several
entanglement purification steps leading to exceedingly
complex architectures. In the direction of improved photodetectors, 
it is worth noticing that single-photon detectors operating at visible 
and near-infrared wavelengths with high detection efficiency and photon-number resolution ability have been developed recently \cite{Lita}, using superconducting transition-edge sensors (TES). 
The current  system detection efficiency at 1556 nm is 95 \%, and may be improved up to the 98-99\% range \cite{Namp}. This would allow such detectors to perform parity measurement at the required quality for the present scheme, obviously without solving the problem on the quantum memory side.

If one wants to distribute entangled
coherent states with realistic resources and rather simple
architectures, one has to consider the limit of small
amplitudes $\alpha \ll 1.$ In this case, errors mainly
reduce to vacuum components and entanglement can be
distributed with high fidelity using postselection.
However, when $\alpha \ll 1,$ entanglement of coherent states reduces to single-photon entanglement. The swapping operations based on linear optics are performed with 50\% of efficiency and quantum repeaters with small-amplitude entangled coherent states do not achieve higher entanglement
generation rates than repeaters based on single-photon entanglement.

Nevertheless, we note that entangled coherent state do not
exploit the full capability of continuous variable
information processing. They are limited to one ebit, like
Bell states. In this sense, the continuous variable scheme studied here
imitates from its premises the discrete variable
architecture. A two mode squeezed state can in principle
reach any value of entanglement, from $0$ to $\infty$
according to the squeezing parameter. For instance, 1.1
ebits have been demonstrated in Ref. \cite{Laurat05} and
recent observations of more than $10$ dB of squeezing
\cite{Schnabel08} would correspond to more than 2.5 ebit if used for
entanglement generation. Such ``EPR-type'' entanglement can
be swapped deterministically using homodyne detection
\cite{Furusawa}. Unfortunately such entanglement cannot be
propagated over long distances. This fragility motivated
the use of the entanglement generation procedure based on
non-local photon subtractions which forms the basis of the
protocol studied in the present paper. Synthesizing remote
entanglement not limited to 1 ebit seems to be an
interesting path to investigate, but, to our knowledge, no protocol has been
proposed so far.

We hope these first results will open further investigations in the frame of quantum repeaters based on continuous variables of light and atoms. \\

We thank M. Afzelius, Q. Glorieux, A. Ourjoumtsev and H. De Riedmatten, R.T. Thew for helpful discussions. This work was supported by the EU through the Integrated Project Qubit Applications and the ICT/FET program Compas, and the Swiss National Foundation through the NCCR Quantum Photonics.


\end{document}